\documentclass[review, number, sort&compress,endfloat]{elsarticle}
\usepackage[utf8]{inputenc}
\usepackage[T1]{fontenc}
\usepackage[american]{babel}
\usepackage{graphicx}
\usepackage{siunitx}

\DeclareSIUnit\torr{Torr}
\DeclareSIUnit\atomic{at.}

\usepackage{booktabs}
\usepackage{tabularx}
\usepackage{miller}
\usepackage[hidelinks]{hyperref}

\usepackage{lineno}

\begin{document}
\title{Relaxation mechanism of GaP grown on 001 Si substrates: influence of defects on the growth of AlGaP layers on GaP/Si templates.}

\author[1]{K.~Pantzas\corref{cor1}}
\ead{konstantinos.pantzas@c2n.upsaclay.fr}

\author[1]{G.~Beaudoin}

\author[1,2]{M.~Bailly}

\author[2]{A.~Martin}

\author[2]{A.~Grisard}

\author[2]{D.~Dolfi}

\author[1]{O.~Mauguin}

\author[1]{L.~Largeau}

\author[1]{I.~Sagnes}

\author[1]{G.~Patriarche}

\cortext[cor1]{Corresponding author}
\address[1]{Université Paris-Saclay, CNRS, Centre de Nanosciences et de Nanotechnologies - C2N, 91120, Palaiseau, France.}

\address[2]{Thales Research \& Technology, F-91120, Palaiseau, France}

\begin{abstract}
    The mechanical stability of commercial GaP/Si templates during thermal annealing and subsequent MOCVD growth of GaP and AlGaP is investigated. Although the GaP layer of the template originally  presents an excellent surface morphology, annealing at high enough temperatures to remove the native oxide prior to growth leads to plastic relaxation, accompanied by a variety of defects, including a dense grid of micro-twins. These micro-twins detrimentally affect GaP and AlGaP layers grown subsequently on the template.
\end{abstract}


\maketitle


\section{Introduction}

The (Al, Ga)P system possesses a significant number of properties highly desirable for nanophotonics integrated on Si. A non-exhaustive list of these properties includes: lattice mismatches of only \SI{0.36}{percent} between GaP and Si \cite{Soga1993}, the smallest among all III-V compounds and Si, and \SI{0.3}{\percent} between AlP and GaP, allowing for the growth of thick, high-quality layers and Bragg mirrors \cite{Beyer2011,IoffeNSM}; a refractive index contrast as large as 0.4 at \SI{700}{\nano\meter} and 0.57 at \SI{500}{\nano\meter}. Only fifteen AlP/GaP pairs are, therefore, required to yield a peak reflectivity of \SI{98}{\percent} for wavelengths between \SIrange{530}{690}{\nano\meter} \cite{Hestroffer2018}; band-gap energies of \SI{2.5}{\electronvolt} and \SI{2.24}{\electronvolt} for AlP and GaP, respectively, making both materials transparent to the green-to-red region of the visible spectrum and allowing active structures to be optically pumped by conventional green lasers; high second-order nonlinearity, Pockels, and piezoelectric effects \cite{IoffeNSM,Aitchinson1997}; high thermal conductivity,  \SI{110}{\watt\per\kelvin\per\meter}, twenty times higher than that of active region material InGaP \cite{Colman2010}, and close to that of Si.

Despite both the low lattice and thermal mismatch between Si and the (Al, Ga)P system, their integration is not straightforward. Indeed, Si crystallizes in the diamond, non-polar structure, while (Al,Ga)P materials crystallize in the polar zinc-blend structure. This difference in polarity leads to the apparition of anti-phase domains (APDs). The presence of such APDs is well-documented in the literature \cite{Beyer2012,Guilleme2016,Lucci2018}, and significant effort has been directed towards their suppression \cite{Beyer2011,Kunert2008}. As a result, GaP on quasi-exact \hkl(001) Si templates are now commercially available. These templates are free of dislocations, stacking-faults and possess an APD-free surface with excellent morphology \cite{NASPWeb}.

In the present contribution, the use of these templates for the growth of GaP and AlGaP using metalorganic chemical vapor deposition (MOCVD) is investigated. Regrowth on these templates typically commences with a high-temperature annealing step in a phosphine ambient to remove the oxide that forms on top of the template. In the conditions developed by the present group of authors for the growth of high-quality, defect-free GaP and $\textnormal{Al}_{0.7}\textnormal{Ga}_{0.3}\textnormal{P}$ layers \cite{XIE2019}, this annealing step as well as the subsequent growth of AlGaP are carried out at \SI{850}{\degreeCelsius}. At this temperature, the lattice mismatch between GaP and Si increases to \SI{0.53}{\percent}. While the mismatch is still low, the critical layer thickness is bound to be exceeded during growth, leading to the relaxation of strain through the generation of misfit dislocations.

The objectives of the present paper are, therefore, to assess the mechanical stability of the GaP/Si template during high-temperature annealing and subsequent growth, and to evaluate the impact of any defects induced by plastic relaxation of misfit strain  on the growth of AlGaP layers. The paper is structured as follows: a succinct description of the GaP/Si templates and the experiments that were carried out is given first. The stability of GaP/Si templates during annealing and the efficiency of the annealing in removing the native oxide are then assessed for an array of temperatures. Finally, the impact that defects induced by the annealing have on the subsequent growth of GaP is discussed.

\section{Experiment}

\begin{figure}
    \centering
    \includegraphics[width = \textwidth]{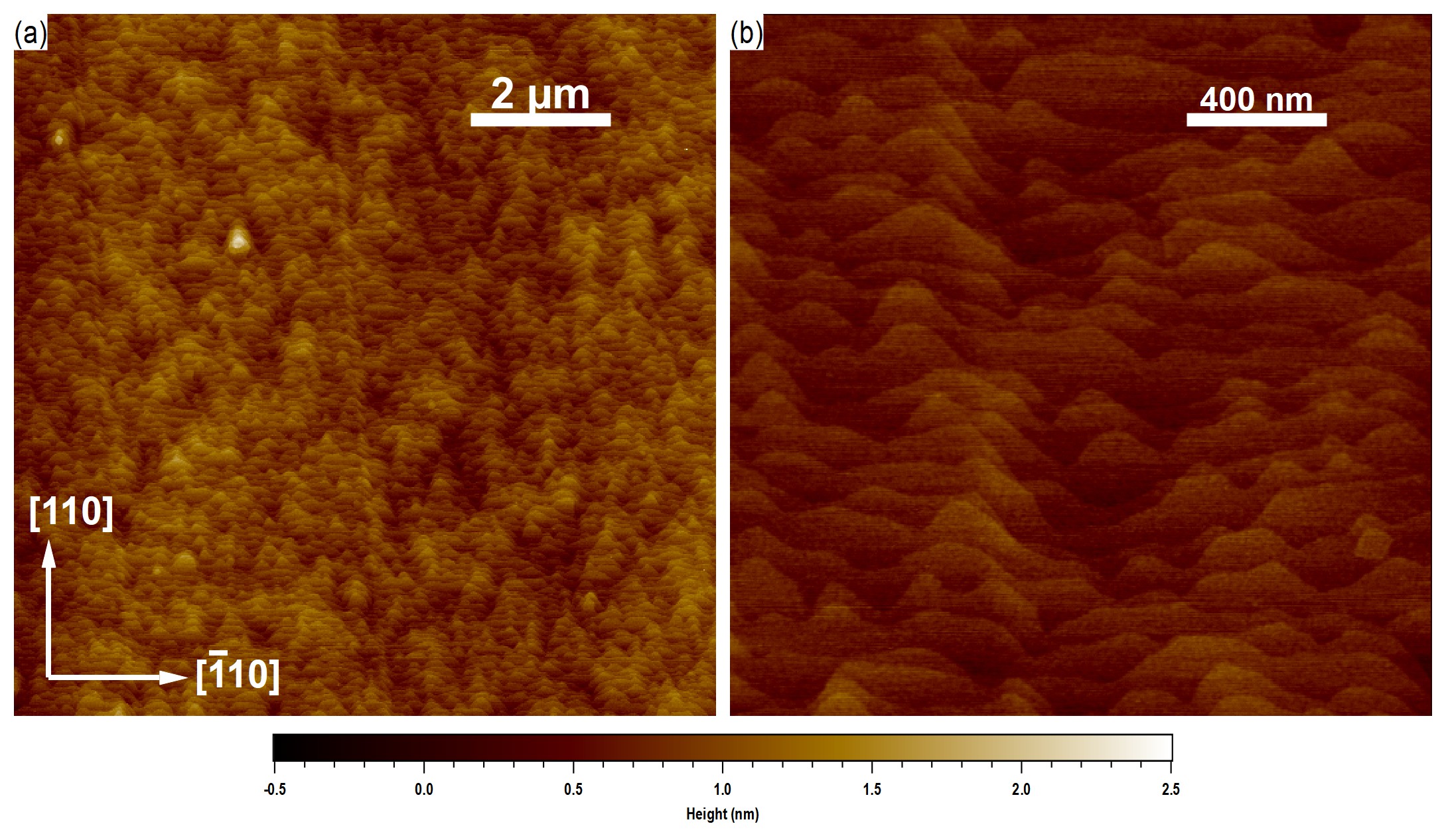}
    \caption{(a) \SI{100}{\micro\meter\squared} and (b) \SI{4}{\micro\meter\squared}  AFM images showing the surface of GaP on a GaP/Si template procured from $\textnormal{NAsP}_\textnormal{III/V}$ GmbH. Step-flow morphology is observed, with the step-flow occurring along the 001 direction. The RMS roughness is \SI{0.2}{\nano\meter}.}
    \label{fig:fig1_afm_ref}
\end{figure}

A \SI{200}{\milli\meter} GaP on Si template was purchased from $\textnormal{NAsP}_\textnormal{III/V}$ GmbH \cite{NASPWeb}. This template consists of \SI{50}{\nano\meter} of GaP on 001 Si. The Si substrate is almost exact, with a small intentional miscut of \SI{0.18}{\degree} in \hkl[110]. According to Reference~\cite{Beyer2011}, this miscut, in conjunction with specific growth conditions, allows one to favor one crystal polarity over the other, resulting  in a GaP/Si template with no observable APDs emerging at the GaP surface. Furthermore, given the small mismatch between GaP and Si - only \SI{0.36}{\percent} at room temperature - the GaP layer is pseudomorphically accommodated on the Si template and, therefore, contains very few dislocations. AFM images of the surface of GaP as received are shown in Figure~\ref{fig:fig1_afm_ref}. The surface exhibits step flow, with a root mean square (RMS) roughness of \SI{0.2}{\nano\meter} and no visible defects in areas as large as \SI{10}{\micro\meter} by \SI{10}{\micro\meter}. 

\begin{figure}
    \centering
    \includegraphics[width=\textwidth]{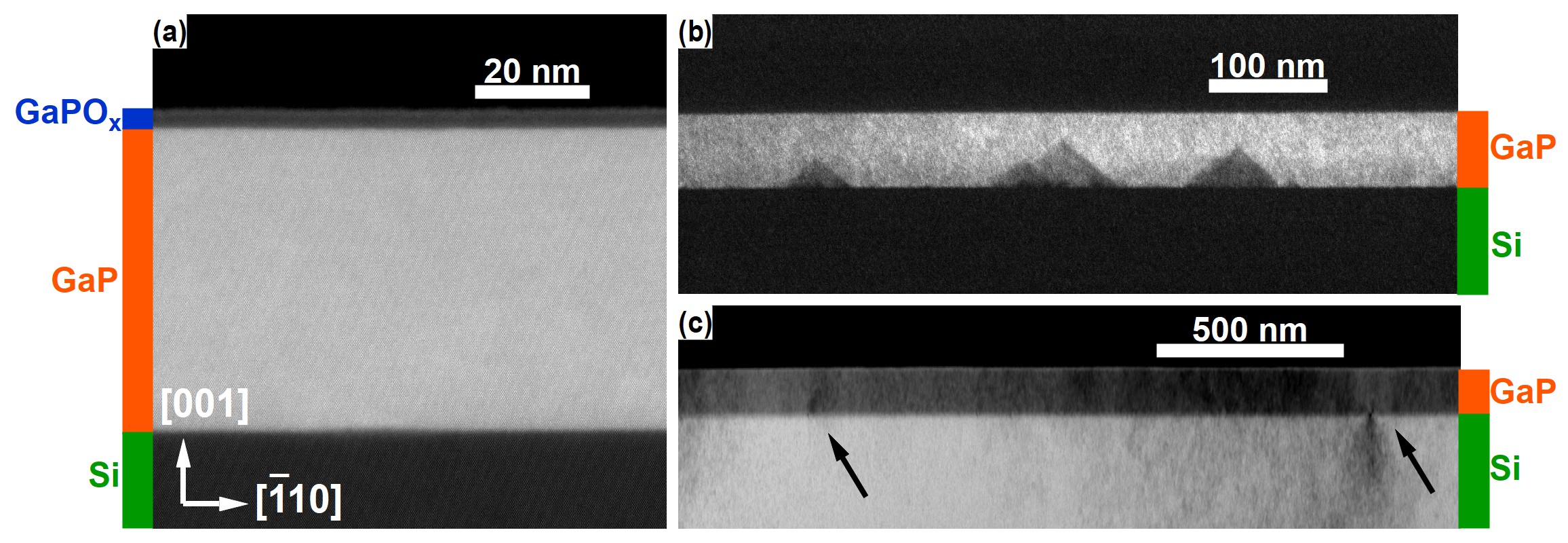}
    \caption{(a)HAADF-STEM image of the as-received GaP/Si template. The image shows that the \SI{50}{\nano\meter} thick GaP layer is covered by a \SI{3}{\nano\meter} thick oxide layer. (b) 002 Dark-field TEM image showing the buried antiphase domains. In all the samples tested none of the domains emerge on the surface. (c) 220 Dark-field TEM image showing dislocations (black arrows) spaced by \SI{1.5}{\micro\meter} along the GaP/Si interface, indicating that despite the low mismatch and the relatively thin GaP layer, plastic relaxation has set on in the GaP/Si template}
    \label{fig:fig2_tem_nasp}
\end{figure}

Cross-sections were prepared for observation in a transmission electron microscope (TEM). The cross-sections for transmission electron microscopy were prepared from some of the samples using Focused Ion Beam (FIB) ion milling and thinning. Prior to FIB ion milling, the sample surface was coated with \SI{50}{\nano\meter} of carbon to protect the surface from the platinum mask deposited used for the ion milling process. Ion milling and thinning were carried out in a FEI SCIOS dual-beam FIB-SEM. Initial etching was performed at \SI{30}{\kilo\electronvolt}, and final polishing was performed at \SI{5}{\kilo\electronvolt}. All TEM cross-sections were observed in an aberration-corrected FEI TITAN 200 TEM-STEM  operating at \SI{200}{\kilo\electronvolt}. A high-angle annular dark-field image of the GaP/Si template is shown in Figure~\ref{fig:fig2_tem_nasp}~(a). The image shows the presence of a \SI{3}{\nano\meter} thick oxide layer on the surface of the \SI{50}{\nano\meter} thick GaP layer. At this scale no defects are present in the GaP layer. Furthermore, 002 Dark-field TEM images of the samples confirm that antiphase domains are indeed buried and do not emerge at the surface of the template - see Figure~\ref{fig:fig2_tem_nasp}~(b). Nevertheless, misfit dislocations can be seen at the GaP/Si interface, as shown in the 220 Dark-field TEM image in Figure~\ref{fig:fig2_tem_nasp}~(c). The grid has an approximate period of \SI{1.5}{\micro\meter}. The presence of this grid of dislocations, indicates the onset of plastic relaxation in the GaP/Si template, despite the small mismatch and relatively thin GaP layer. No threading dislocations are visible in the lamella, however. Given that the cross-sections are \SI{25}{\micro\meter} long by \SI{80}{\nano\meter} wide, the threading-dislocation density is less than \SI{5e7}{\per\centi\meter\squared}.

The \SI{200}{\milli\meter} template was diced into \SI{50}{\milli\meter} wafers using laser cutting. Annealing and growth was carried out in a Veeco D180 Turbodisk reactor. The annealing experiments were carried out at three different annealing temperatures were used: \SI{645}{\degreeCelsius}, \SI{745}{\degreeCelsius}, and \SI{845}{\degreeCelsius}, for \SI{30}{\minute} in a phosphine ambient. The lowest temperature is close to the one disclosed in Reference~\cite{Beyer2011}, discussing the fabrication of the templates. The highest temperature is similar to one used for the obtained high quality GaP and AlGaP in Reference~\cite{XIE2019}. In addition to the annealing experiments, two samples were grown on the GaP/Si templates using MOCVD: a \SI{200}{\nano\meter} GaP layer to test GaP growth, and, separately, the following structure to test AlGaP growth: a nominally \SI{325}{\nano\meter} thick GaP buffer followed by two periods consisting of a \SI{47}{\nano\meter} thick AlGaP layer, containing \SI{73}{\atomic\percent} aluminum, and a \SI{64}{\nano\meter} thick GaP layer. Additionally, the \SI{80}{\nano\meter} of the GaP buffer contain five \SI{1}{\nano\meter} thick AlP markers, spaced every \SI{16}{\nano\meter}. Both samples were grown at a reactor temperature of \SI{845}{\degreeCelsius} and a reactor pressure of \SI{20}{\torr}, using  trimethylaluminum, trimethylgallium, and phopshine as precursors to elementary aluminum, gallium, and phosphorous, respectively.

\section{Results}

\begin{figure}
    \centering
    \includegraphics[width=\textwidth]{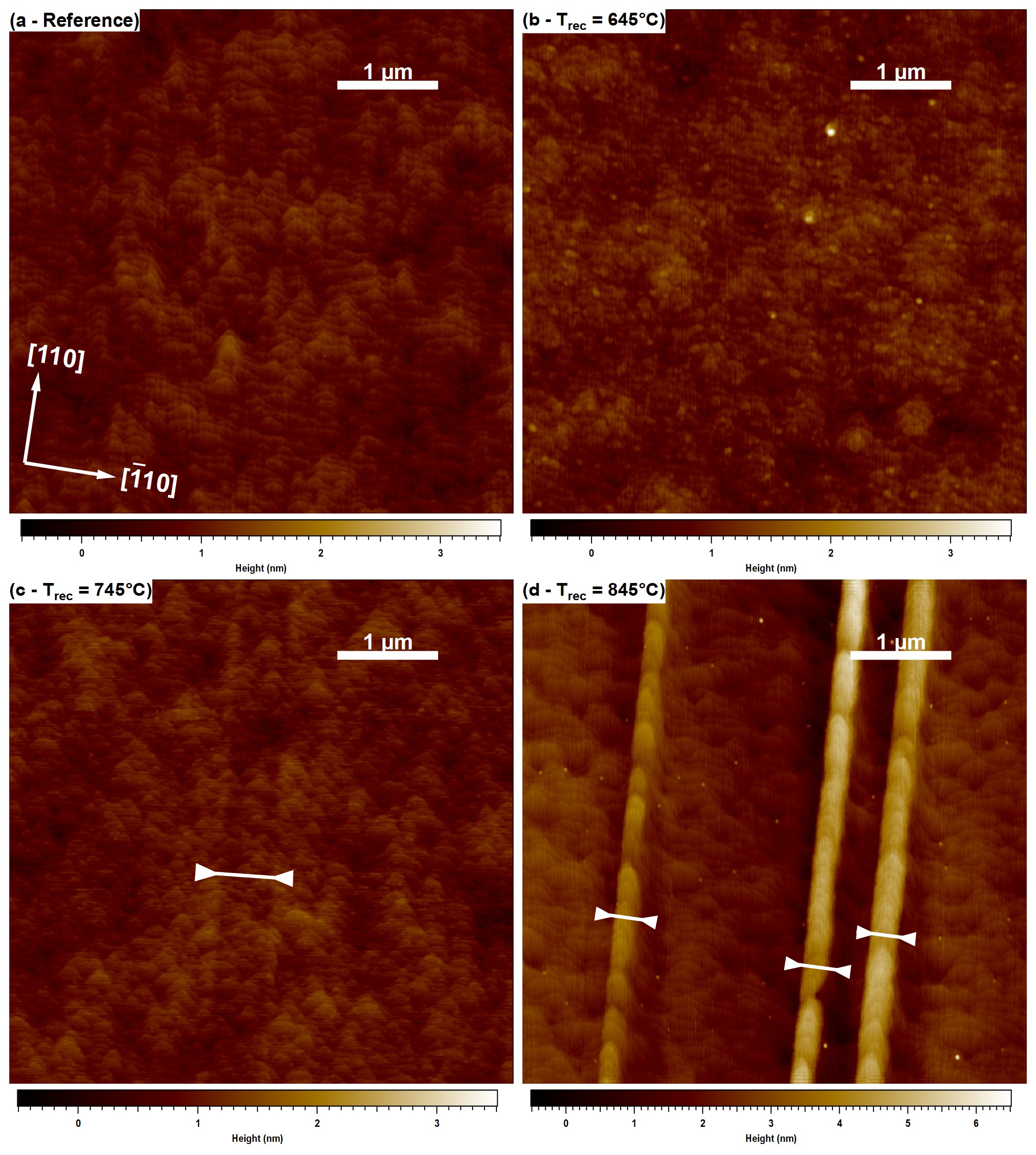}
    \caption{\SI{25}{\micro\meter\squared} AFM images of the surface of the GaP/Si template (a) before, and after annealing at (b) \SI{645}{\degreeCelsius}, (c) \SI{745}{\degreeCelsius}, and (d) \SI{845}{\degreeCelsius} for \SI{30}{\minute}. At \SI{645}{\degreeCelsius} the atomic steps are preserved, but the surface is riddled with small pockets of oxide that was not removed. At higher annealing temperatures, these pockets disappear and the step-flow morphology is preserved. At \SI{745}{\degreeCelsius} and \SI{845}{\degreeCelsius}, however, bands that are on average \SI{375}{\nano\meter} wide appear to rise above from the background. These are particularly visible in the sample annealed at \SI{845}{\degreeCelsius}. At this temperature the elevation of these bands is \SI{5}{\nano\meter}. These bands are indicative of plastic relaxation taking place in the GaP layer of the template during high temperature annealing.}
    \label{fig:fig3_afm_anneal}
\end{figure}

The four AFM images in Figure~\ref{fig:fig3_afm_anneal} show the evolution of the surface morphology of the GaP/Si template as a function of temperature. At the lowest annealing temperature, \SI{645}{\degreeCelsius}, the GaP layer preserves it's surface morphology, exhibiting step-flow and no visible defects. Nevertheless, the surface still exhibits spheres of oxide, indicating that the annealing temperature was not sufficient to completely remove the \SI{3}{\nano\meter} oxide present at the surface of the template. Annealing for an hour at the same temperature did not yield any different results (not shown here). At higher temperatures, the oxide is completely removed and the step-flow morphology is still present, but the surface also exhibits stripes that are between \SI{250}{\nano\meter} and \SI{500}{\nano\meter} in width. At the highest annealing temperature, \SI{845}{\degreeCelsius}, these stripes are more markedly present, creating a \SI{3}{\nano\meter} high step. Additionally, several dislocations emerge at the surface of GaP, proof that the GaP layer plastically relaxes during the thermal annealing. The estimated threading-dislocation density is \SI{2e8}{\per\centi\meter\squared}, at least an order of magnitude higher than that present in the template initially.

\begin{figure}
    \centering
    \includegraphics[width=\textwidth]{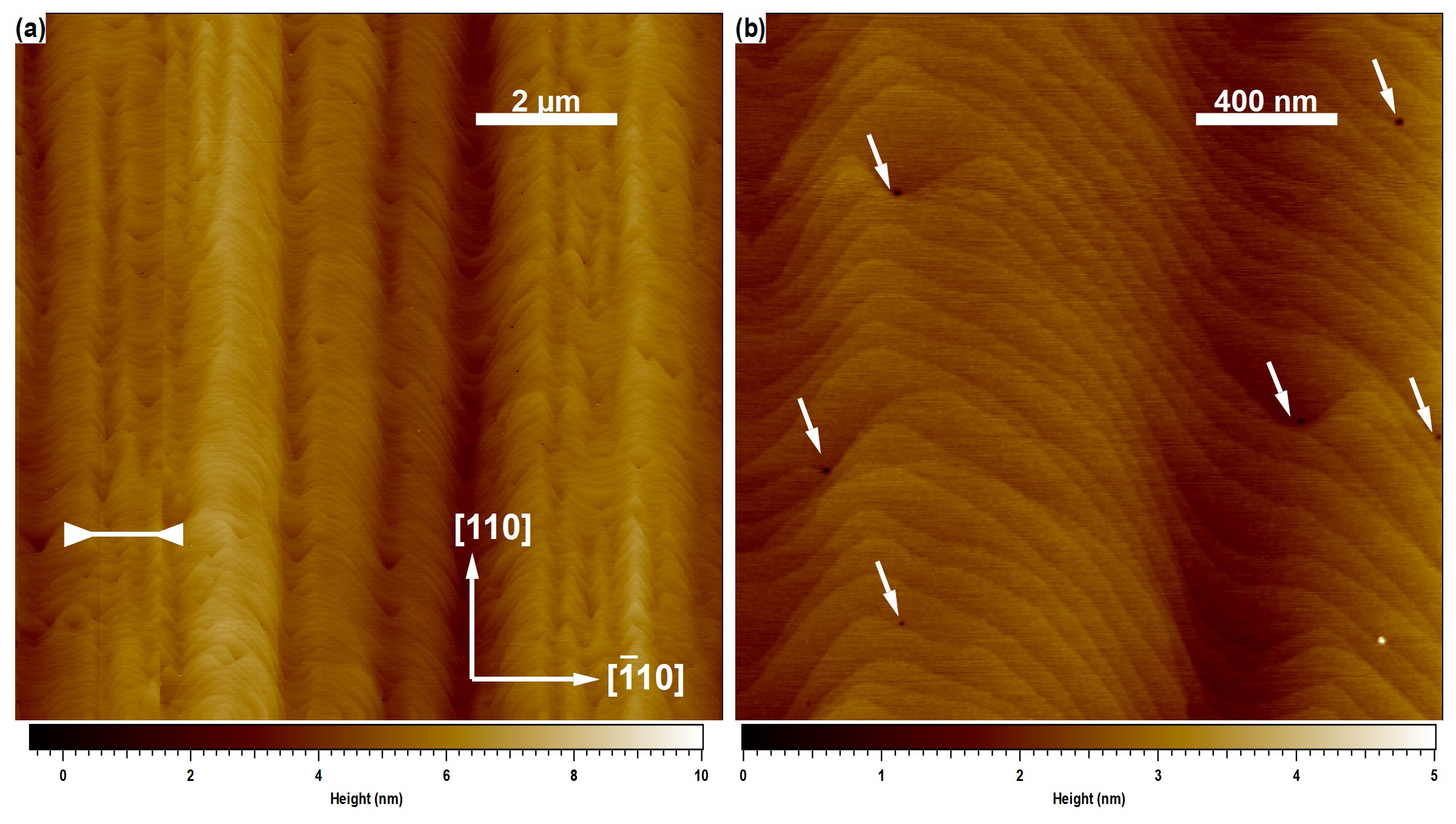}
    \caption{(a) \SI{100}{\micro\meter\squared} and (b) \SI{4}{\micro\meter\squared} AFM images of the GaP surface after the growth of a \SI{200}{\nano\meter} of GaP on the GaP/Si template using MOCVD. Although step flow is still observed, the surface undulates along the \hkl<010> direction. This undulation is due to crosshatch. Crosshatch in GaP has been shown to occur anisotropically, more dense along the \hkl<010> direction than the \hkl<100> direction \cite{Lucci2018}. The surface is dotted by small pits where threading dislocations emerge, pointed out by white arrows in image (b). The threading dislocation density is estimated at \SI{1e8}{\per\centi\meter\squared}, four orders of magnitude higher than that of the initial template. In addition to the emergence pits and the crosshatch, at least one \SI{500}{\nano\meter} wide band delimited by two lines is pointed out by arrows in image (a), indicating that the bands that appear during annealing persist throughout subsequent growth of GaP.}
    \label{fig:fig4_afm_gap}
\end{figure}

The defects observed on the annealed samples persist in the \SI{200}{\nano\meter} GaP layer grown on the GaP template, as shown the AFM images of the sample surface in Figure~\ref{fig:fig4_afm_gap}. The threading dislocation density, estimated from the density of pits in the AFM images, is \SI{2e8}{\per\centi\meter\squared}, no different than that observed for the sample at \SI{845}{\degreeCelsius}. The GaP surface is also rougher owing to the onset of cross-hatch, with an RMS roughness of \SI{0.9}{\nano\meter}, although step-flow is still preserved.

\begin{figure}
    \centering
    \includegraphics[width=\textwidth]{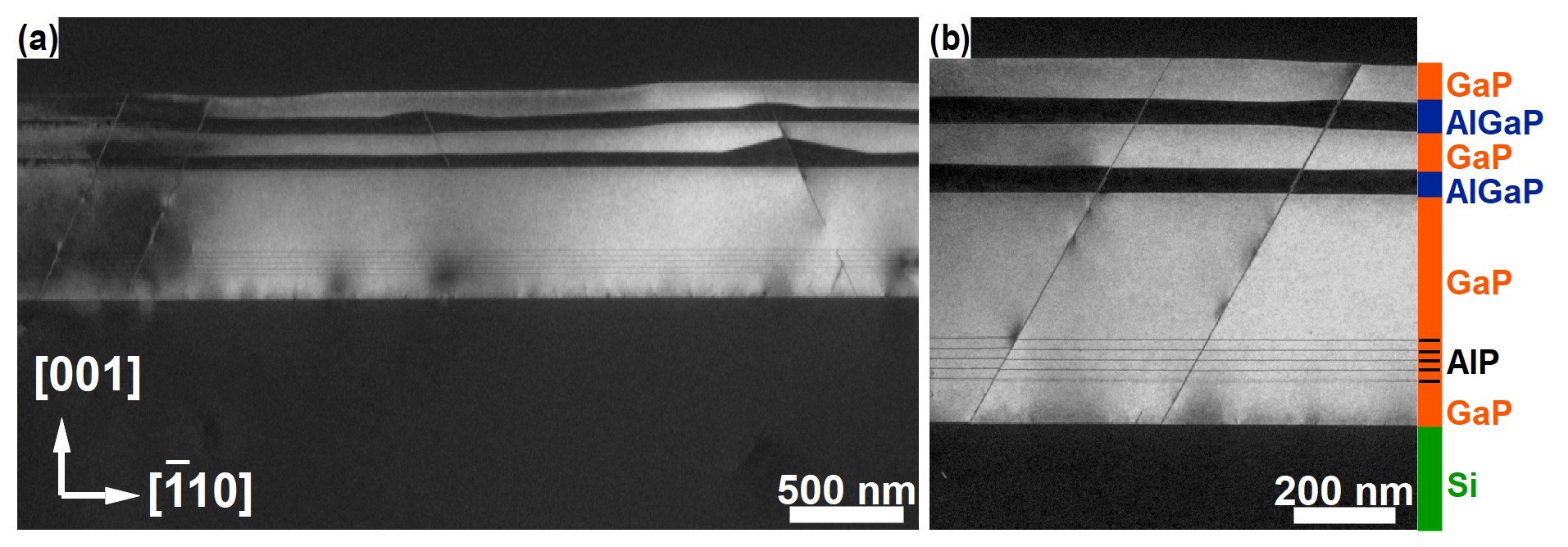}
    \caption{(a) 220 dark-field TEM image of the AlGaP/GaP structure grown on the GaP template. A dense grid of misfit dislocation is present at the GaP/Si interface, indicating that the structure is fully plastically relaxed. In addition to the dislocation grid, micro-twins appear in the structure. These often appear in pairs spaced by \SI{365}{\nano\meter}, as revealed in (b) showing a 220 dark-field TEM image close up of one such pair. These pairs are linked to . These micro-twins induce faceting in AlGaP, significantly roughening the GaP/AlGaP interfaces. This faceting is highly detrimental to the optical properties of AlGaP/GaP pairs. As shown in Figure~\ref{fig:fig3_afm_anneal}, these micro-twins already appear during the annealing of the GaP/Si template.}
    \label{fig:fig5_tem_algap}
\end{figure}

The defects observed in the GaP sample significantly perturb the growth of AlGaP, as revealed by the dark-field TEM images of the AlGaP sample, shown in Figure~\ref{fig:fig5_tem_algap}. These images reveal that the stripes observed already during annealing are bordered by pairs of micro-twins. When these cross the AlGaP layer, they induce severe faceting, mainly \hkl[112] and \hkl[114] planes. This faceting is mainly visible at the AlGaP/GaP interface. Subsequent growth of GaP tends to smooth this roughness, as seen at the second GaP/AlGaP interface in Figure~\ref{fig:fig5_tem_algap}, but the surface becomes again rougher during the growth of the second AlGaP layer. The RMS roughness at the surface is \SI{5}{\nano\meter}, more than an order of magnitude higher than the initial roughness of the template.

\begin{figure}
    \centering
    \includegraphics[width=\textwidth]{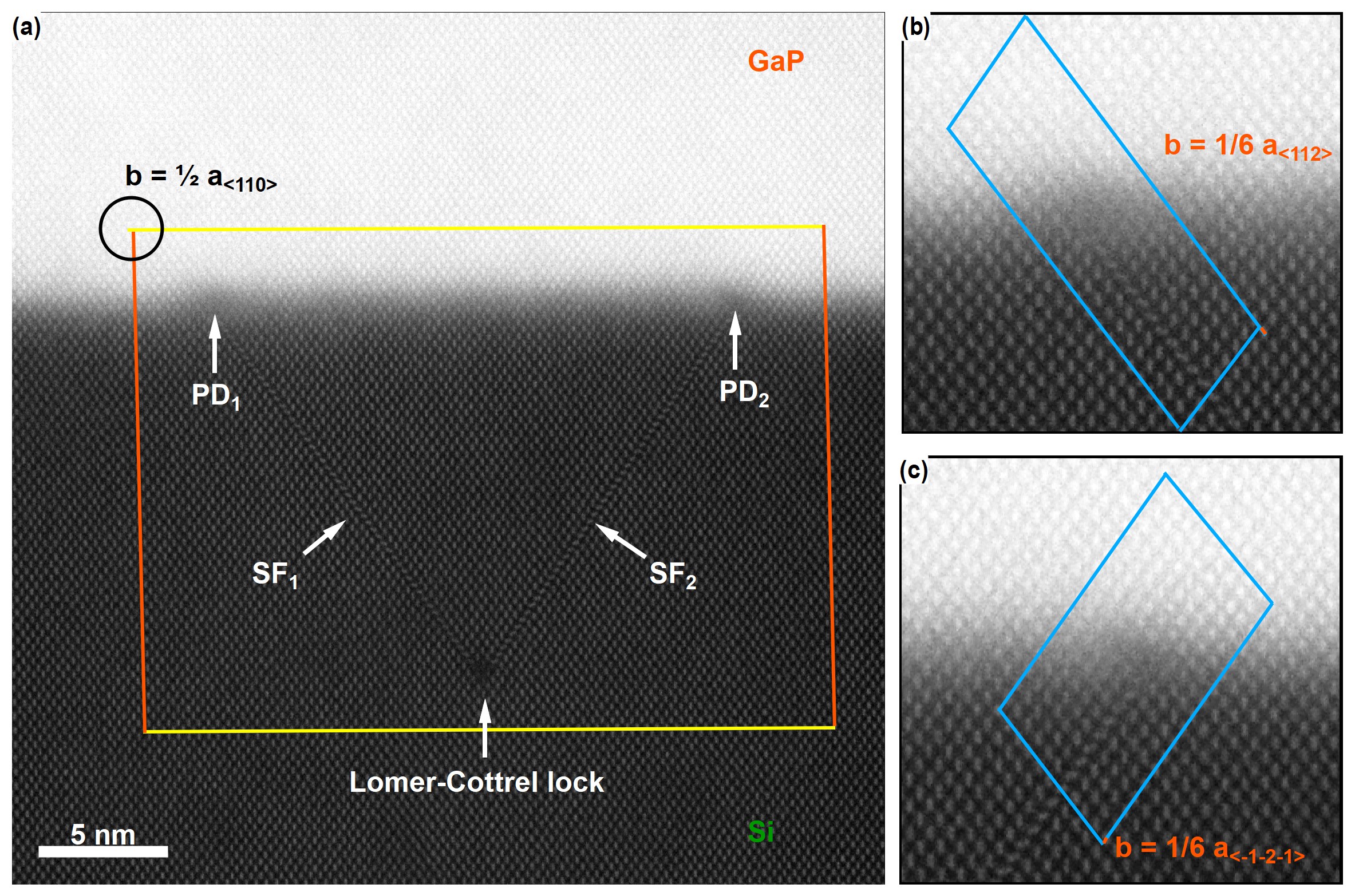}
    \caption{(a) HAADF-STEM image of a the GaP/Si interface in the AlGaP sample showing a stair-rod dislocation. The stair-rod forms at the junction of the stacking faults (SF$_1$ and SF$_2$) that extend from two partial dislocations (PD$_1$ and PD$_2$) at the GaP/Si interface and meet in the Si substrate, forming a sessile Lomer-Cottrel dislocation. (b,c) Insets showing magnified portions of the HAADF-STEM image in (a), centered around PD$_1$ and PD$_2$, and their Burgers' circuits.}
    \label{fig:fig6_stairrod}
\end{figure}

Finally, the GaP/Si interface now exhibits a dense misfit-dislocation grid, with an approximate period of \SI{30}{\nano\meter}. Upon close inspection, this dislocation grid is composed up to \SI{60}{\percent} of edge dislocations, and \SI{40}{\percent} partial dislocations. Half of the partial dislocations at the interface combine with a Lomer-Cottrel dislocation in Si to form a stair-rod dislocation \cite{Hull2011}. This type of defect is more common in highly mismatched systems, such as Ge/Si that has a lattice mismatch of \SI{4}{\percent}, an order of magnitude higher than GaP/Si \cite{Arroyo2019}, and to the extent of our knowledge has not been reported for lowly mismatched systems. The stair-rod itself appears as an inverted triangle at the GaP/Si interface. One such defect is shown in Figure~\ref{fig:fig6_stairrod}~(a). Two apexes of the the triangle are partial misfit dislocations at the GaP/Si interface. Tracing the Burgers' circuits of the partial dislocations - Figures~\ref{fig:fig6_stairrod}~(b) and (c) - reveals that their Burgers' vector is $\mathbf{b} = \frac{1}{6}\mathbf{a}_{112}$, in keeping with theory. Stacking faults extend from them and slip along \hkl[111] planes in Si until they meet at about \SI{30}{\nano\meter} from the GaP/Si interface. There they combine to form the Lomer-Cottrel dislocation, that has a Burgers' vector of $\mathbf{b} = \frac{1}{2}\mathbf{a}_{110}$ and is sessile.

\section{Discussion}

Both AFM and TEM results show that the \SI{50}{\nano\meter} GaP layer in the template does not contain any emerging APDs, and presents a smooth surface with a step-flow morphology and a roughness of only \SI{0.2}{\nano\meter}. Nevertheless, the TEM images reveal that even at this thickness, which is lower than the critical thickness for plastic relaxation of GaP, reported at \SI{90}{\nano\meter} \cite{Soga1996}, a loose misfit-dislocation grid is already present at the GaP/Si interface. In the literature, this has been associated with thermal-stress induced relaxation of GaP during the cooling down phase of the GaP/Si template following GaP growth at temperatures higher  than \SI{400}{\degreeCelsius}. The reported threading dislocation densities in optimized conditions were of the order of \SI{1e7}{\per\centi\meter\squared} \cite{Tachikawa1990,Olson1986}, in agreement with the estimate obtained from TEM on the reference sample.

In two samples grown by MOCVD, the total GaP thickness is more than \SI{250}{\nano\meter}, exceeding the critical thickness. In the literature full plastic relaxation of misfit strain that is due to lattice mismatch between GaP and Si is not achieved until a thickness of \SI{200}{\nano\meter} \cite{Olson1986,Soga1996,Emmer2017}. At the reported thickness thickness the samples are, therefore, fully plastically relaxed, and, accordingly, exhibit dislocation densities in the \SI{1e8}{\per\centi\meter\squared} range. The plastic relaxation of the strain induced to mismatch is further supported by the appearance of Lomer-Cottrel locks in Si. Indeed, the appearance of misfit dislocations in the Si substrate has been previously associated with the relaxation of strain that is due to the lattice mismatch, whereas thermal stress tends to generate defects at the GaP/Si interface or in GaP \cite{Tachikawa1990}. Although the RMS roughness of GaP sample increases, it still remains within reasonable limits, i.e. less than \SI{1}{\nano\meter}. This roughness could be improved by increasing the V/III ratio and/or the reactor pressure.

The most cumbersome defects, however, are the micro-twins that already appear during high-temperature annealing of the GaP/Si template. These defects induce heavy faceting in the AlGaP layer, rendering these layers inappropriate for the growth of AlGaP/GaP Bragg mirrors. The micro-twins reported here are denser when the annealing temperature is increased. This result is contrary to what has been observed in the case of GaP grown on \SI{4}{\degree} off-cut Si using migration-enhanced molecular-beam epitaxy \cite{Yamane2010}. Indeed, in that reference, micro-twins are eliminated by implementing an annealing step after the initial nucleation of a \SI{20}{\nano\meter} GaP layer. It is unclear whether the miscut is solely responsible for this difference, or whether the micro-twins could be annihilated by further tweaking of the growth and annealing conditions on the current templates. Alternatively, growth could be performed in a temperature range closer to what has been reported by NAsP$_\textnormal{III/V}$ Gmbh., i.e. between \SI{450}{\degreeCelsius} and \SI{600}{\degreeCelsius} \cite{Beyer2011}, using precursors that are better suited to low-temperature growth, such as triethylgallium and tri-tert-butylphosphine. Nevertheless, incorporation of high amounts of aluminum at low temperatures may prove challenging.

\section{Conclusion}

Despite excellent surface properties and the absence of emerging APDs, the GaP/Si templates used in the present contribution are revealed not to be compatible with the growth conditions developed of high-quality AlGaP and GaP growth on GaP substrates. Indeed, the templates are unstable at high temperatures, plastically relaxing to form micro-twins that detrimentally affects the growth of AlGaP in particular. Furthermore, the oxide layer present at the surface of the templates is not readily removed through annealing in a phosphine ambient at lower temperatures. A combination of wet etching to remove the oxide and growth with alternative metalorganic precursors that are better suited to low-temperature growth may be the only solution towards growth of high quality GaP and AlGaP on these templates.

\section*{Acknowledgements}
The authors would like acknowledge J. Nagle, A. De Rossi, and S. Combrié for their support and fruitful discussions. The authors thankfully acknowledge funding from the CNRS Renatech network and the ANR Labex TEMPOS (ANR-10-EQPX-0050).

\bibliographystyle{unsrtnat}

\end{document}